\documentclass[8.5pt,twoside,twocolumn]{article}
\usepackage[super,sort&compress,comma]{natbib}
\usepackage{mhchem}
\usepackage{times,mathptmx}
\usepackage{sectsty}
\usepackage{balance}

\usepackage{graphicx} %eps figures can be used instead
\usepackage{lastpage}
\usepackage[format=plain,justification=raggedright,singlelinecheck=false,font=small,labelfont=bf,labelsep=space]{caption}
\usepackage{fancyhdr}
\usepackage{color}
\newcommand{\esi}{ESI\dag}

\newcommand{\ib}[1]{{\color{black}#1}}

\begin{document}

\thispagestyle{plain}
\fancypagestyle{plain}{
% \fancyhead[L]{\includegraphics[height=8pt]{headers/LH}}
% \fancyhead[C]{\hspace{-1cm}\includegraphics[height=20pt]{headers/CH}}
% \fancyhead[R]{\includegraphics[height=10pt]{headers/RH}\vspace{-0.2cm}}
\renewcommand{\headrulewidth}{1pt}}
\renewcommand{\thefootnote}{\fnsymbol{footnote}}
\renewcommand\footnoterule{\vspace*{1pt}%
\hrule width 3.4in height 0.4pt \vspace*{5pt}}
\setcounter{secnumdepth}{5}

\makeatletter
\def\subsubsection{\@startsection{subsubsection}{3}{10pt}{-1.25ex plus -1ex minus -.1ex}{0ex plus 0ex}{\normalsize\bf}}
\def\paragraph{\@startsection{paragraph}{4}{10pt}{-1.25ex plus -1ex minus -.1ex}{0ex plus 0ex}{\normalsize\textit}}
\renewcommand\@biblabel[1]{#1}
\renewcommand\@makefntext[1]%
{\noindent\makebox[0pt][r]{\@thefnmark\,}#1}
\makeatother
\renewcommand{\figurename}{\small{Fig.}~}
\sectionfont{\large}
\subsectionfont{\normalsize}

\fancyfoot{}
% \fancyfoot[LO,RE]{\vspace{-7pt}\includegraphics[height=9pt]{headers/LF}}
% \fancyfoot[CO]{\vspace{-7.2pt}\hspace{12.2cm}\includegraphics{headers/RF}}
% \fancyfoot[CE]{\vspace{-7.5pt}\hspace{-13.5cm}\includegraphics{headers/RF}}
\fancyfoot[RO]{\footnotesize{\sffamily{1--\pageref{LastPage} ~\textbar  \hspace{2pt}\thepage}}}
\fancyfoot[LE]{\footnotesize{\sffamily{\thepage~\textbar\hspace{3.45cm} 1--\pageref{LastPage}}}}
\fancyhead{}
\renewcommand{\headrulewidth}{1pt}
\renewcommand{\footrulewidth}{1pt}
\setlength{\arrayrulewidth}{1pt}
\setlength{\columnsep}{6.5mm}
\setlength\bibsep{1pt}

\newcommand{\alt}{\raisebox{-0.3ex}{$\stackrel{<}{\sim}$}}
\newcommand{\agt}{\raisebox{-0.3ex}{$\stackrel{>}{\sim}$}}
\graphicspath{{ }{.}{./calc/}{./kai/}}

\twocolumn[
  \begin{@twocolumnfalse}
\noindent\LARGE{\textbf{Concurrent conductance and transition voltage spectroscopy study of 
scanning tunneling microscopy vacuum junctions. Does it unravel new physics? $^{\ast}$\dag
% \footnote{This article is a companion paper and is intended to be read in conjunction with the article DOI: 10.1039/C4RA04651 published in RSC Adv.~by H.~J.~W.~Zandvliet \emph{et al}.}
}}
\vspace{0.6cm}

\noindent\large{\textbf{Ioan B\^aldea \textit{$^{a\ddag}$}
}}\vspace{0.5cm}

\noindent \textbf{\small{Published: RSC Advances 2014, {\bf 4}, 33257-33261; DOI: 10.1039/C4RA04648J}}
\vspace{0.6cm}

\noindent 
\normalsize{Abstract:\\
We show that the conductance ($G$) and transition voltage ($V_t$) spectroscopy data for standard vacuum nanogaps reported in the preceding paper by Sotthewes \emph{et al} cannot be understood within the framework of the existing theories/models whatever realistic effects are incorporates. However, if we include an additional (``ghost'') current, the trends of the dependencies of $G$ and $V_t$ on the nanogap size $d$ can be explained. The ghost current is very small. Therefore, effects related to the ghost current can only be revealed at larger $d$'s, where the ghost current becomes comparable to or overcomes the tunneling current. Although we are not able to unravel the origin of this ghost contribution, we can and do refer to experimental $G$- and $V_t$-data reported for molecular junctions, which exhibit similarities to the presently considered vacuum nanojunctions. Analyzing notable difficulties related to the regime assigned as transport via hopping, we speculate that a ghost contribution may also exist in molecular junctions.

$ $ \\  

}
\vspace{0.5cm}
 \end{@twocolumnfalse}
  ]

%Footnotes
%Please use \dag to cite the ESI in the main text of the article.
%If you article does not have ESI please remove the the \dag symbol from the title and the above footnotetext.

\footnotetext{\textit{$^{a}$~Theoretische Chemie, Universit\"at Heidelberg, Im Neuenheimer Feld 229, D-69120 Heidelberg, Germany.}}
\footnotetext{\ddag~E-mail: ioan.baldea@pci.uni-heidelberg.de.
Also at National Institute for Lasers, Plasmas, and Radiation Physics, Institute of Space Sciences,
Bucharest, Romania}
\footnotetext{$^\ast$~This article is a companion paper and is intended to be read in conjunction with the article DOI: 10.1039/C4RA04651 published in RSC Adv.~by H.~J.~W.~Zandvliet \emph{et al}.}
\footnotetext{\dag~Electronic supplementary information (ESI) available. See DOI:10.1039/C4RA04648J}
\section{Introduction}
\label{sec:intro}
A long standing challenge in the field of molecular electronics is whether 
the measured currents are really mediated by the active molecules intentionally 
embedded in the nanodevice under consideration
or are (also) significantly affected, \emph{e.g.}, by impurities or 
defects.\cite{kushmerick:09,Reed:11}
From this perspective, the transport through vacuum nanodevices seems to be privileged, 
because it is hard to imagine how could the measured current $I \equiv I_{measured}$  be plagued by undesired 
``ghost'' channels giving rise to a current $I_{ghost}$ 
overimposed on the current $I_{tunnel}$ due to the tunneling across the vacuum nanogap. 
However strange it might be, after ruling out all causes we could conceive, 
we have to conclude that the challenging experimental results 
on the transition voltage $V_t$ spectroscopy (TVS) 
\cite{Beebe:06,kushmerick:09}
reported in the preceding paper \cite{harold-letter}
cannot be explained without assuming a ``ghost'' contribution. 
The origin of such a 
transport channel contributing to the transport in standard 
vacuo remains the big mystery, which the present work cannot unravel. Still, what we can 
presently do is to point out to
some similarities with results reported for molecular junctions 
\cite{Choi:08,Choi:10,Tao:10} and suggest that a similar contribution
can also be present there.
%%%%%%%%%%%%%%%%%%%%%%%%%%%%%%%%%%%%%%%%%%%%%%%%%%%%%%%%%%%%%%%%%%%%%%%%%%%%%
\section{Theoretical details}
\label{sec:theory}
%%%%%%%%%%%%%%%%%%%%%%%%%%%%%%%%%%%%%%%%%%%%%%%%%%%%%%%%%%%%%%%%%%%%%%%%%%%%%
The tunneling current through a vacuum nanogap can be 
determined by exact numerical solution of the Schr\"odinger equation
by using a tunneling barrier
$\phi_{B}$ that includes contribution from the applied bias $V$, electrodes' 
(STM substrate $s$ and tip $t$) work functions 
$\Phi_{s,t}$, and image charges $\phi_{im}$ 
\cite{Sommerfeld:33,Baldea:2012c}
%%%%%%%%%%%%%%%%%%%%%%%%%%%%%%%%%%%%%%%%%%%%%%%%%%%%%%%%%%%%%%%%%%%%%%%%% 
\begin{eqnarray}
\label{eq-phiB} 
\phi_{B}(z; V) & = &  \left(\Phi_{t} + \Phi_{s}\right)/2  + \phi_{im}(z) + V_{b}(z) ,\\
\label{eq-V_b}
V_{b}(z) & = & \left(\Phi_{s} - \Phi_{t} - e V\right) z/d .
\end{eqnarray}
%%%%%%%%%%%%%%%%%%%%%%%%%%%%%%%%%%%%%%%%%%%%%%%%%%%%%%%%%%%%%%%%%%%%%%%%% 
The ``bare'' rectangular barrier due the electrodes' work function 
is tilted by the applied bias $V$ and reduced by (classical) image charges
$\phi_{im}(z)$.   For two planar electrodes, the exact 
expression of $\phi_{im}$ \cite{Sommerfeld:33} 
can be used to numerically find the exact transmission coefficient
\cite{Baldea:2012e,Baldea:2012f}.

\begin{figure}[htb]
$ $\\[0ex]
\centerline{\includegraphics[width=0.4\textwidth,angle=0]{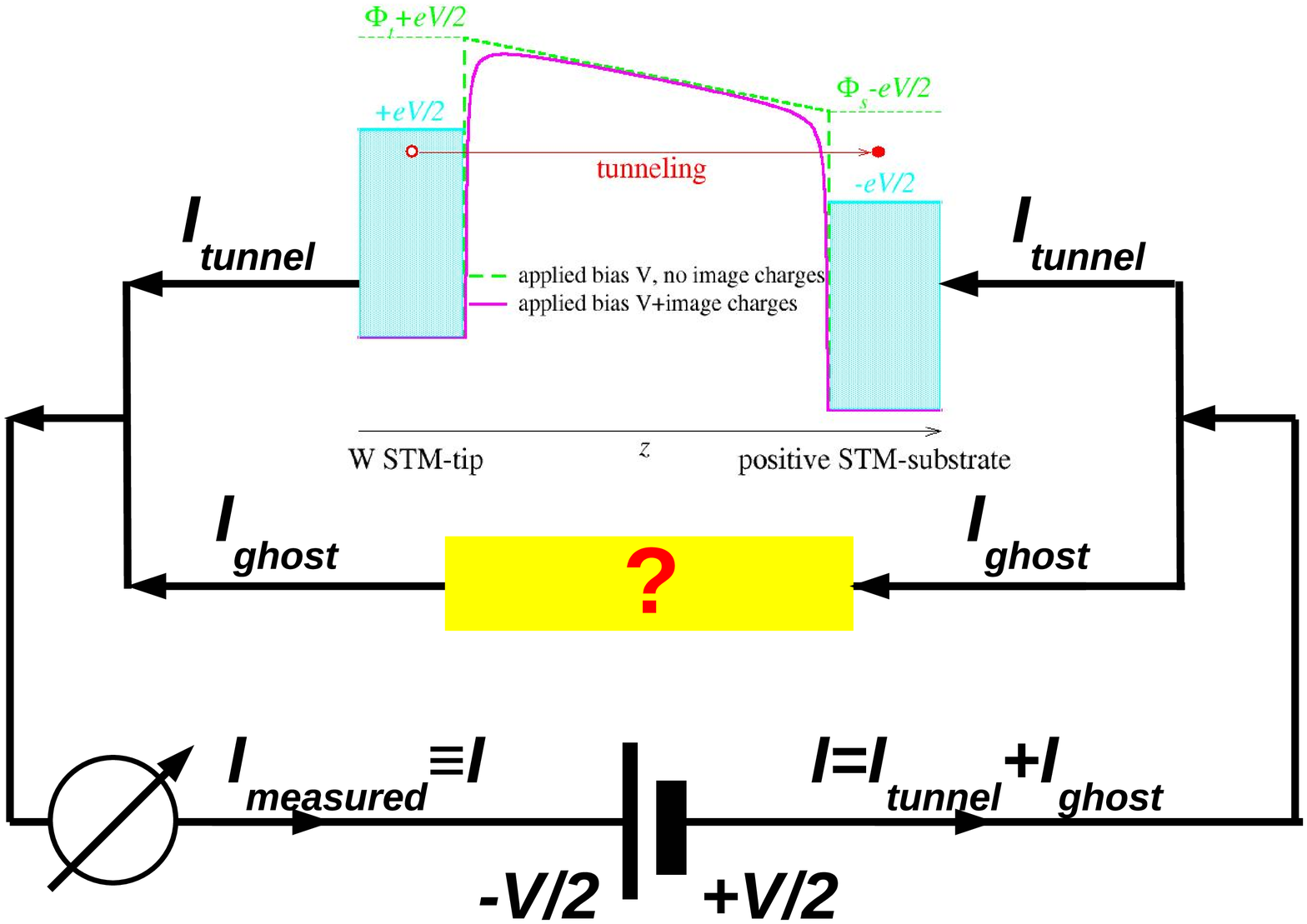}}
$ $\\[0ex]
\caption{Schematic representation of the transport scenario proposed in the present paper.}
\label{fig:cartoon}
\end{figure}

In the discussion that follows we will assume that, besides the current $I_{tunnel}$  
due to electron 
tunneling through a metal-vacuum-metal junction in a scanning tunneling microscope (STM) setup, 
the measured current also comprises a ``{ghost}'' current $I_{ghost}$
(\emph{cf.}~\ref{fig:cartoon})
%%%%%%%%%%%%%%%%%%%%%%%%%%%%%%%%%%%%%%%%%%%%%%%%%%%%%%%%%%%%%%%%%%%%%%%%%%%%%%
\begin{equation}
\label{eq-I}
I \equiv I_{measured} = I_{tunnel} + I_{ghost} .
\end{equation}
%%%%%%%%%%%%%%%%%%%%%%%%%%%%%%%%%%%%%%%%%%%%%%%%%%%%%%%%%%%%%%%%%%%%%%%%%%%%%%
%%%%%%%%%%%%%%%%%%%%%%%%%%%%%%%%%%%%%%%%%%%%%%%%%%%%%%%%%%%%%%%%%%%%%%%%% 
%%%%%%%%%%%%%%%%%%%%%%%%%%%%%%%%%%%%%%%%%%%%%%%%%%%%%%%%%%%%%%%%%%%%%%%

For simplicity, we will assume that the latter represents an 
Ohmic contribution, that is, 
\begin{equation}
\label{eq-I-ghost}
I_{ghost} = \frac{V}{\rho d} .
\end{equation}
Here $V$ is the applied bias between the STM tip and substrate, which 
are spatially separated by a vacuum nanogap of size $d$, and $\rho$ stands for the {ghost}
(average) resistivity. 

Below, we will focus our attention on the Ohmic conductance $G$ and the 
transition voltage $V_t$.
$V_t$ is defined as the bias at the minimum of the Fowler-Nordheim quantity 
$\ln (I/V^2)$ \cite{Beebe:06}, or, equivalently, by the point where the
differential conductance is two times larger than the pseudo-Ohmic conductance
\cite{Baldea:2012b}.
%%%%%%%%%%%%%%%%%%%%%%%%%%%%%%%%%%%%%%%%%%%%%%%%%%%%%%%%%%%%%%%%%%%%%%%%%
The interplay between the tunneling and ghost contributions 
of eqn~(\ref{eq-I}) will be characterized by the dimensionless 
quantity $\overline{\rho}$ related to the {ghost} resistivity $\rho$ 
%%%%%%%%%%%%%%%%%%%%%%%%%%%%%%%%%%%%%%%%%%%%%%%%%%%%%%%%%%%%%%%%%%%%%%%%% 
\begin{equation}
\label{eq-rho}
\overline{\rho} \simeq 1.618 \times 10^{4}\,{\rho}\ \Omega^{-1} \cdot \mbox{m}^{-1} .
\end{equation}
%%%%%%%%%%%%%%%%%%%%%%%%%%%%%%%%%%%%%%%%%%%%%%%%%%%%%%%%%%%%%%%%%%%%%%%%% 
%%%%%%%%%%%%%%%%%%%%%%%%%%%%%%%%%%%%%%%%%%%%%%%%%%%%%%%%%%%%%%%%%%%%%%%%%%%%%
\section{Results}
\label{sec:results}
%%%%%%%%%%%%%%%%%%%%%%%%%%%%%%%%%%%%%%%%%%%%%%%%%%%%%%%%%%%%%%%%%%%%%%%%%%%%%
\subsection{Attempting to incorporate realistic effects into the tunneling barrier}
\label{sec:wo-ghost}
%%%%%%%%%%%%%%%%%%%%%%%%%%%%%%%%%%%%%%%%%%%%%%%%%%%%%%%%%%%%%%%%%%%%%%%%%%%%%
The results for $V_t$ presented in
\ib{Fig.~1 and 5} of ref.~\citenum{harold-letter}, showing that calculations based on
Simmons-type approximation, similar to those done earlier \cite{Huisman:09,Molen:11},
cannot explain the $d$-dependence of the transition voltage $V_t$ found in experiment
(\emph{cf}.~\ib{Fig.~4A and 4B} of ref.~\citenum{harold-letter}) may not be surprising; 
shortcomings of the Simmons approach
to transition voltage spectroscopy are amply documented 
\cite{Baldea:2012c,Baldea:2012e,Baldea:2012h}. 

In the first part of the present analysis, we will show 
that the tunneling transport model
cannot explain the experimental $V_t$-data even when
the numerical exact solution of the 
Schr\"odinger equation is employed (this is the case for all results presented here), 
and even attempting to make the barrier model more realistic.
This is shown by the results presented in 
Fig.~S1
% Fig.~\ref{fig:vt-wo-ghost-realistic} 
of the {\esi}, 
which emerged from the attempt
to make the tunneling barrier as realistic as possible.
(Throughout, labels S refer to the {\esi}.)

\begin{figure}[htb]
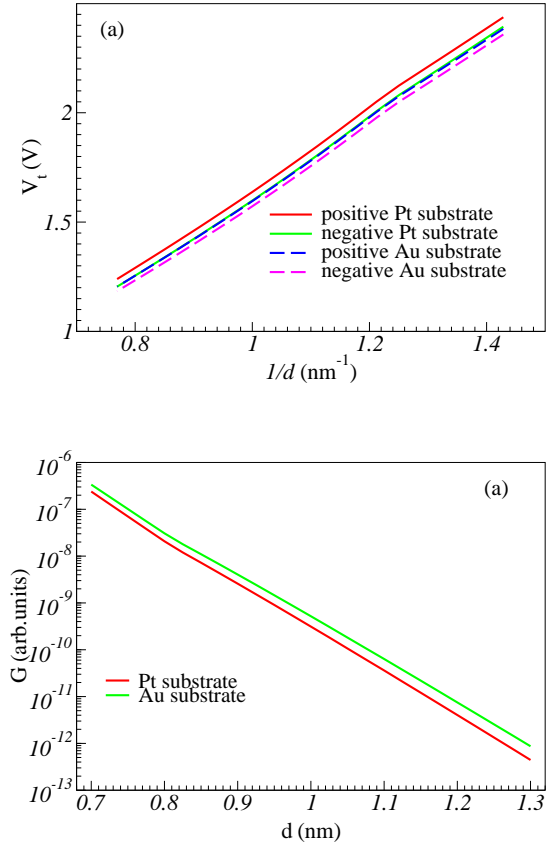

$ $\\[4ex]
\centerline{
\includegraphics[width=0.4\textwidth,angle=0]{fig_vt_Pt_Au_no_ghost.eps}}
$ $\\[1ex]
\centerline{
\includegraphics[width=0.4\textwidth,angle=0]{fig_g_Pt_Au_no_ghost.eps}}
% $ $\\[1ex]
\caption{(a) Transition voltage $V_t$ and (b) Ohmic conductance 
as a function of the nanogap size $d$ 
for STM-tip of tungsten ($\Phi_{W}=4.55$\,eV) and 
substrates of Pt 
($\Phi_{Pt} = 5.65$\,eV) and Au ($\Phi_{Au} = 5.2$\,eV). Image charge effects are 
included using the exact classical interaction \cite{Baldea:2012e,Baldea:2012f,Baldea:2012h}, 
which is cutoff close to electrodes using the procedure described elsewhere
\cite{Baldea:2013b}.}
\label{fig:vt-g-wo-ghost}
\end{figure}

Fig.~\ref{fig:vt-g-wo-ghost}a shows the
effect of the difference between the STM-tip and substrate 
work functions ($\Phi_{t} \to \Phi_{W} = 4.55$\,eV versus
$\Phi_{s} \to \Phi_{Pt} = 5.65$\,eV or $\Phi_{Au}=5.2$\,eV, respectively, \emph{cf.}~eqn~(\ref{eq-phiB}).
This may be important, because it yields an extra Volta-type field of a strength 
$\sim \vert \Phi_s - \Phi_t\vert/(ed) \sim 1$\,V/nm comparable to the applied field $V/d$
\cite{Sommerfeld:33,Gundlach:66,Baldea:2012f}. As seen in Fig.~\ref{fig:vt-g-wo-ghost}a, 
the asymmetry $\Phi_{s} \neq \Phi_{t}$ semi-quantitatively explains the small 
difference $V_{t+} \neq \vert V_{t-}\vert $ 
between positive and negative biases for Pt substrates. As noted \cite{harold-letter}, the Shockley surface 
state, not included in the present calculations, represents a plausible source for the considerably 
larger asymmetry $V_{t+} \neq \vert V_{t-}\vert $ for Au (111) substrates. 
% Fig.~\ref{fig:vt-wo-ghost-realistic}a 
Fig.~S1a demonstrates that image effects, which may be important at smaller sizes 
\cite{Gundlach:66}, are not essential for the $d$-values used in experiments \cite{harold-letter}.
The same figure indicates that the potential profile has no significant impact on $V_t$
% (cf.~Fig.~\ref{fig:vt-wo-ghost-realistic}a).
(\emph{cf.}~Fig.~S1a).
Instead of well defined $d$- and $\Phi$-values, it might be more realistic to 
consider statistical distributions of these quantities. These distributions may better 
model STM-tips without a well defined facet and polycrystalline Pt substrates.
But, as visible in 
% Fig.~\ref{fig:vt-wo-ghost-realistic}b and Fig.~\ref{fig:vt-wo-ghost-realistic}c, 
Fig.~S1b and S1c, 
these effects do not improve the agreement with experiment.
%%%%%%%%%%%%%%%%%%%%%%%%%%%%%%%%%%%%%%%%%%%%%%%%%%%%%%%%%%%%%%%%%%%%%%%%% 
%%%%%%%%%%%%%%%%%%%%%%%%%%%%%%%%%%%%%%%%%%%%%%%%%%%%%%%%%%%%%%%%%%%%%%%

Fig.~\ref{fig:vt-g-wo-ghost}b depicts results for 
the Ohmic conductance $G$, a quantity that also plays an
important part in the subsequent analysis. Whatever the effects 
considered (\emph{cf.}~Fig S2), it 
does not notably deviate from the well-known exponential decay. 
This is also at variance with the experimental $G$-data, which
evolve to a plateau at larger $d$-values 
(\emph{cf.}~Figure \ib{4C} of ref.~\citenum{harold-letter}).
%%%%%%%%%%%%%%%%%%%%%%%%%%%%%%%%%%%%%%%%%%%%%%%%%%%%%%%%%%%%%%%%%%%%%%%
%%%%%%%%%%%%%%%%%%%%%%%%%%%%%%%%%%%%%%%%%%%%%%%%%%%%%%%%%%%%%%%%%%%%%%%%%%%%%
\subsection{Impact of a ghost current}
\label{sec:with-ghost}
\begin{figure}[htb]
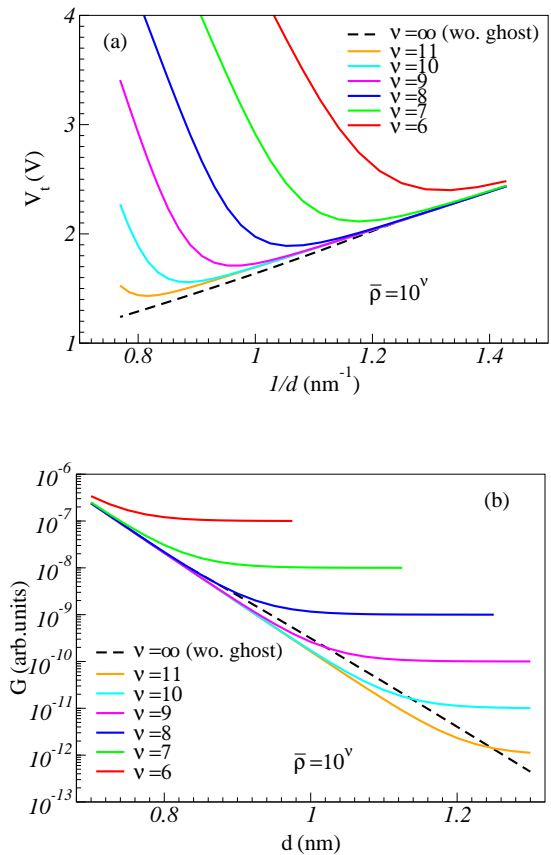

$ $\\[4ex]
\centerline{
\includegraphics[width=0.4\textwidth,angle=0]{fig_vt_positive_Pt_with_ghost.eps}}
$ $\\[1ex]
\centerline{
\includegraphics[width=0.4\textwidth,angle=0]{fig_g_Pt_with_ghost.eps}
}
$ $\\[1ex]
\caption{(a) Transition voltage $V_t$ and (b) Ohmic conductance $G$ as a function of the nanogap size $d$ 
for STM-tip of tungsten ($\Phi_{W}=4.55$\,eV) and substrates of Pt at positive biases 
($\Phi_{Pt} = 5.65$\,eV) in the presence of a ghost channel of dimensionless resistivity
$\overline{\rho} = 10^{\nu}$. $\nu$-values are given in the legend. 
Image charge effects are 
included using the exact classical interaction \cite{Baldea:2012e,Baldea:2012f,Baldea:2012h}, 
which is cutoff close to electrodes using the procedure described elsewhere
\cite{Baldea:2013b}.}
\label{fig:vt-g-ghost}
\end{figure}

%%%%%%%%%%%%%%%%%%%%%%%%%%%%%%%%%%%%%%%%%%%%%%%%%%%%%%%%%%%%%%%%%%%%%%%%%%%%%
Let us now assume that a ghost current [eqn~(\ref{eq-I-ghost})] exists, and 
examine its impact on $G$ and $V_t$. Results for $G$ and $V_t$ at various values of the
parameter $\overline{\rho}$ are presented in Fig.~\ref{fig:vt-g-ghost}a and Fig.~\ref{fig:vt-g-ghost}b.
The trend observed in these figures is compatible with the experimental behavior.

In the plateau regime ($d \simeq 1.12$\,nm), experiments found
$G_{Pt} \simeq 5 \times 10^{-9}$\,S for platinum substrates and $G_{Au} \simeq 5 \times 10^{-10}$\,S
for gold substrates
(\emph{cf}.~Figure \ib{4C} of ref.~\citenum{harold-letter}).
A reliable estimate for the transverse area vacuum nanogap depends, 
amongst others, on the actual tip shape and cannot be given \cite{harold-private}.
So, let us assume a tip radius $r =1; 10; 100$\,nm. Inserting the above values
in eqn~(\ref{eq-rho}), we get $\overline{\rho} \sim 10^5; 10^7; 10^9$ for Au and 
$\overline{\rho} \sim 10^4; 10^6; 10^8$ for Pt. These estimates  
fall in the $\overline{\rho}$-range 
employed in Fig.~\ref{fig:vt-g-ghost}, wherein the increase of $V_t$ with $1/d$ 
predicted by the conventional theory switches to the decrease 
observed in experiments \cite{harold-letter}.
So, a ghost current may be at the origin of both Ohmic conductances saturating at larger $d$'s 
and transition voltages that decrease with $1/d$. We have checked that this trend is 
not affected if the realistic effects discussed in the preceding subsection and the {\esi}
are included in calculations.
%%%%%%%%%%%%%%%%%%%%%%%%%%%%%%%%%%%%%%%%%%%%%%%%%%%%%%%%%%%%%%%%%%%%%%%%% 
%%%%%%%%%%%%%%%%%%%%%%%%%%%%%%%%%%%%%%%%%%%%%%%%%%%%%%%%%%%%%%%%%%%%%%%

\begin{figure}[htb]
$ $\\[4ex]
\centerline{\includegraphics[width=0.4\textwidth,angle=0]{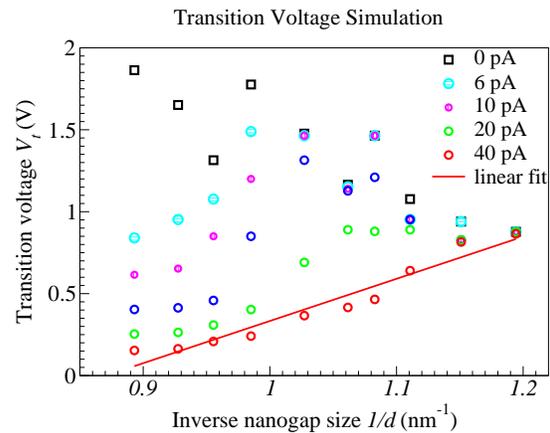}}
$ $\\[1ex]
\caption{By subtracting larger and larger ghost currents 
($I_{ghost}$-values at $V=1$\,V specified in the legend) from the measured current, 
the measured $d$-dependence of $V_t$ (black squares, same as the black circles in 
Fig.~4b of ref.~\cite{harold-letter}) switches from decreasing  
to increasing with $1/d$. The latter dependence is similar to that depicted in 
Fig.~\ref{fig:vt-g-wo-ghost}a expected without ghost contribution.
(Courtesy of the authors of ref.~\citenum{harold-letter}, who allowed
to use here their experimental data.)
}
\label{fig:vt-simulation}
\end{figure}

% \subsection{Simulation}\label{sec:simulation}
%%%%%%%%%%%%%%%%%%%%%%%%%%%%%%%%%%%%%%%%%%%%%%%%%%%%%%%%%%%%%%%%%%%%%%%
As a supplementary test related to the most unexpected experimental 
finding of ref.~\citenum{harold-letter} (namely, a $V_t$ decreasing with $1/d$),
we present further results in Fig.~\ref{fig:vt-simulation}.
The black squares shown there are experimental data $V_t$ vs.~$1/d$. 
The other symbols depicted in Fig.~\ref{fig:vt-simulation} correspond to $V_t$ obtained 
by subtracted from the experimental $I-V$ curves a (ghost) current
proportional to $V$ [\emph{cf.}~eqn~(\ref{eq-I-ghost})], 
which has been gradually increased
from cyan-magenta-blue-green to red circles 
($I_{ghost}$-values are indicated in the legend). 
In spite of certain scattering in the data, the trend is pretty clear: the $V_t$ data 
switch from a $V_t$ decreasing with $1/d$ to a $V_t$ increasing with $1/d$.
The differences between the red circles and the 
red line (linear fit of the red circles) of  Fig.~\ref{fig:vt-simulation} 
are within the experimental errors.
We assign the last curve (red circles) as a situation where
the ghost current has been eliminated. So, what remained is the genuine 
tunneling current, and the corresponding 
curve (red points) $V_t\ vs.\ 1/d$ 
behaves like the curves found  
in Fig.~\ref{fig:vt-g-wo-ghost}a within the conventional theory, 
wherein merely the tunneling current has been accounted for.
The red circles correspond to a ghost resistance $R_{ghost} \approx 10$\,G$\Omega$.
Assuming again tip radii $r=1; 10; 100$\,nm, we find 
$\overline{\rho} \sim 0.4 \times 10^6; 0.4 \times 10^8; 0.4 \times 10^{10}$, which is 
compatible with the range covered in Fig.~\ref{fig:vt-g-ghost}.
%%%%%%%%%%%%%%%%%%%%%%%%%%%%%%%%%%%%%%%%%%%%%%%%%%%%%%%%%%%%%%%%%%%%%%%%% 
%%%%%%%%%%%%%%%%%%%%%%%%%%%%%%%%%%%%%%%%%%%%%%%%%%%%%%%%%%%%%%%%%%%%%%%

To end this section, we note that neither the $d$- nor the $V$-dependence of the ghost current 
assumed in eqn~(\ref{eq-I-ghost}) is critical for the behavior depicted in Fig.~\ref{fig:vt-g-ghost}.
To check the impact of the $1/d$-dependence of eqn~(\ref{eq-I-ghost}), 
we performed calculations by using a $d$-independent ghost resistance
and found no significant modification.
Likewise, to check the impact of the $V$-dependence, instead of 
the contribution of eqn~(\ref{eq-I-ghost}), we added a small constant term 
$\tau$ to the transmission across the tunneling barrier
[\emph{i.e.}, using $\mathcal{T}(E_z; V) \to \mathcal{T}(E_z; V) + \tau$ in 
% eqn~(\ref{eq-J-3d}).
eqn~(S1)].
Results obtained in this way, which are presented in 
% Fig.~\ref{fig:vt-g-tau-ghost}, 
Fig.~S3, 
do not 
qualitatively differ from those of Fig.~\ref{fig:vt-g-ghost}.
Notice that this procedure --- which amounts to consider a ``ghost'' 
transmission channel, a pseudo-diffusion or hopping --- yields via 
% eqn~(\ref{eq-J-3d}) 
eqn~(S1) 
a nonlinear ghost current. To summarize the results based on further
functional dependencies studied, we can state that the qualitative behavior
depicted in Fig.~\ref{fig:vt-g-ghost} holds provided that the 
$d$- and $V$-dependencies of the ghost current are much weaker than 
the corresponding (exponential) dependencies of the tunneling current.
%%%%%%%%%%%%%%%%%%%%%%%%%%%%%%%%%%%%%%%%%%%%%%%%%%%%%%%%%%%%%%%%%%%%%%%%%%%%%
\section{Discussion}
\label{sec:discussion}
%%%%%%%%%%%%%%%%%%%%%%%%%%%%%%%%%%%%%%%%%%%%%%%%%%%%%%%%%%%%%%%%%%%%%%%%%%%%%
We have shown above that the experimental data for 
electron transport in STM vacuum junctions can be rationalized by postulating
the existence of a ghost current. What could be the origin of the ghost current?
A imperfect electric insulation cannot come in question \cite{harold-private}. 
The vacuum in experiments is very good ($< 3 \times 10^{-11}$\,mbar);
therefore it is very unlikely that a
conduction channel through the vacuum junction is responsible for the ghost 
currents needed in Fig.~\ref{fig:vt-g-ghost} and Fig.~\ref{fig:vt-simulation}.
To rule out a ghost current flowing through the electronic circuit, the experimentalists 
have checked that for large tip-sample separation ($\sim 2$\,nm) 
the measured current is negligible \cite{harold-private}.
A small DC offset voltage of the $I-V$ converter gives an unavoidable, small contribution 
($\sim 1 - 2$\,pA) to the measured current, which should be independent on the bias; 
however, any residual offset current has been subtracted before processing the data
\cite{harold-private}. Therefore, this possibility should also be ruled out.

So, we are unable to unravel the origin of 
the ghost current postulated above (even, see below) in a
standard vacuum nanogap. Still, we believe that this issue may be related to 
pieces of experimental work on molecular junctions 
of two independent groups. 
Some time ago, Frisbie's group first reported a crossover between 
two different transport regimes in molecular junctions employing 
the conducting probe atomic force microscopy (CP-AFM) setup
\cite{Choi:08,Choi:10}. For shorter molecules, the resistance
$R$ was found temperature ($T$) independent and increasing
exponentially with increasing length ($d$), in contrast to longer molecules
characterized by an Arrhenius $T$-dependence $R\propto \exp(E_a/k_B T)$ 
and a linear $d$-dependence. 
Extracting information on single-molecules using transport data on CP-AFM
junctions \cite{Choi:08,Choi:10}, comprising many ($\sim 100$) molecules, 
might be not straightforward. 
A subsequent work of Tao's group \cite{Tao:10} on single-molecule STM-junctions
confirmed the crossover between two transport regimes with the same 
signatures noted above \cite{Choi:08,Choi:10}. 
Fully in light with current understanding of charge transport, 
these $d$- and $T$-dependencies have been taken as clear
evidence of charge transport via tunneling in shorter molecules 
and via hopping in longer molecules \cite{Choi:08,Choi:10,Tao:10}. 

While by no means intending to challenge that, with increasing $d$,
tunneling becomes less effective and hopping increasingly important,
for reasons exposed below we believe that
this is not the whole issue related to the transport data of 
refs.~\citenum{Choi:08}, \citenum{Choi:10}, and \citenum{Tao:10}.
Particularly problematic to us appear the very large values of 
the activation energies extracted from experiments, an aspect on which 
ref.~\citenum{Choi:10} already drew attention. 
As discussed in the {\esi}, ref.~\citenum{Choi:10} drastically overestimated
the reorganization energies $\lambda$, 
because the contributions of all intramolecular vibrations were included in calculations;
in reality, only very low frequency modes, which can be thermally activated, 
can contribute \cite{Baldea:2014a}. Even though, the (over)estimated $\lambda$-values
remain several times smaller than those needed to explain 
the large experimental activation energies $E_{a}$. 
The measured values $E_{a}\simeq 0.54 - 0.62$\,eV \cite{Choi:10}
and $E_{a}\simeq 0.58 - 0.58$\,eV \cite{Tao:10} would require 
enormous reorganization energies $\lambda > 2$\,eV 
% [cf.~eqn~(\ref{eq-lambda}]
[\emph{cf.}~eqn~(S5)]
hardly compatible with known (including author's) results 
of quantum chemical calculations.

\begin{figure}[htb]
$ $\\[4ex]
\centerline{\includegraphics[width=0.4\textwidth,angle=0]{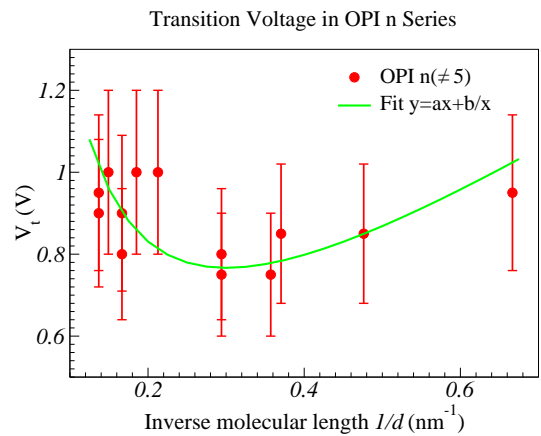}}
$ $\\[0ex]
\caption{The transition voltage of the OPI series exhibits 
a dependence on the molecular size ($n$)compatible 
to that found in Fig.~\ref{fig:vt-g-ghost}a. The points represent experimental results 
(\emph{cf.}~Table S1 of ref.~\citenum{Choi:10}). 
The error bars correspond to relative standard deviations in $V_t$ of $\sim 20$\%, 
as typically found in experiments \cite{Beebe:06,Guo:11}.
(Size $n = 5$ has been eliminated
since it seems to correspond to a special crossover situation \cite{Choi:08}.)
}
\label{fig:choi10}
\end{figure}

The $d$-dependence of the resistance $R$ 
of the aforementioned experimental works on molecular junctions 
\cite{Choi:08,Choi:10,Tao:10} is similar to the presently
considered vacuum nanojunctions 
(compare Figure 2A of ref.~\citenum{Choi:08}, Figure 6b of ref.~\citenum{Choi:10},
and Figure 3 of ref.~\citenum{Tao:10} with Figure \ib{4C} 
of ref.~\citenum{harold-letter} and the present Fig.~\ref{fig:vt-g-ghost}b).
This represents one important fact for the present analysis, 
but not the only one. In addition to results for $R$ vs.~$d$, 
ref.~\citenum{Choi:08} reported results on $V_t$ for oligophenyleneimines (OPIs)
of various sizes. Interestingly, the trend visible in Fig.~\ref{fig:choi10},
wherein these results are depicted,
resembles the trend of Fig.~\ref{fig:vt-g-ghost}a.
So, it is tempting to interpret the similarity between 
the $d$-dependence of Fig.~\ref{fig:choi10} and
Fig.~\ref{fig:vt-g-ghost}a as suggesting that a ghost current may also exist 
in molecular junctions.    
%%%%%%%%%%%%%%%%%%%%%%%%%%%%%%%%%%%%%%%%%%%%%%%%%%%%%%%%%%%%%%%%%%%%%%%%% 
%%%%%%%%%%%%%%%%%%%%%%%%%%%%%%%%%%%%%%%%%%%%%%%%%%%%%%%%%%%%%%%%%%%%%%%
%%%%%%%%%%%%%%%%%%%%%%%%%%%%%%%%%%%%%%%%%%%%%%%%%%%%%%%%%%%%%%%%%%%%%%%%%%%%%
\section{Conclusion}
\label{sec:conclusion}
%%%%%%%%%%%%%%%%%%%%%%%%%%%%%%%%%%%%%%%%%%%%%%%%%%%%%%%%%%%%%%%%%%%%%%%%%%%%%
TVS was initially proposed as a tool of extracting 
the energy offset of the dominant frontier orbital relative to 
electrode's Fermi energy in molecular junctions \cite{Beebe:06},
and a series of studies \cite{Baldea:2012a,Baldea:2012b,Baldea:2012g,Baldea:2013b}
demonstrates that it holds its promise. The fact that $V_t$ 
can be used to quantify a significant nonlinear
transport \cite{Baldea:2012b} makes TVS a method of interest
even beyond molecular electronics, and ref.~\citenum{harold-letter}
as well as the present paper have illustrate its utility 
for vacuum nanojunctions.

We have shown that
the conductance and TVS data for a standard vacuum nanogap 
reported in ref.~\citenum{harold-letter}
cannot be understood within the framework
of the existing theories/models whatever realistic effects are incorporates. 
However, if we include an additional (``ghost'') current, 
the experimental trends $G$ vs.~$d$ and $V_t$ vs.~$d$ 
exhibited by the Ohmic conductance and transition voltage can be explained. 
The ghost current is very small; it could not have played a significant
role in ref.~\cite{Molen:11} because at the very small nanogap sizes 
employed there ($d<0.4$\,nm \cite{Molen:11})
the tunneling current is too large, and it masks a possible ghost contribution.
Effects related to the ghost current
can only be revealed at larger nanogap sizes $d$ (like those of ref.~\cite{harold-letter}), 
where the ghost current overcomes the tunneling current. 
This is the reason why we believe that incorporating electrodes' atomistic structure 
(\emph{e.g.}, within approaches based on the density functional theory \cite{Hou:12,Hou:13})
will not change the conclusions of the present study, which is mostly based on the classical barrier picture for charge transport via tunneling \emph{in vacuo}. 
Such details can be quite relevant for TVS in \emph{other} cases: \emph{e.g.}, for \emph{vacuum} nanogaps of sizes smaller than those ($d > 0.8$\,nm) presently considered \cite{Hou:12}, or \emph{molecular} junctions \cite{Hou:13}, where through-space processes are negligible.

The analysis of existing studies 
on size-dependence of $G$ and $V_t$ in molecular junctions
have indicated that a ghost current may also be present in those 
nanosystems, which could have a significant role in 
properly assessing and characterizing 
the transport mechanism (e.g.~hopping). 
From this perspective, investigating the impact of temperature 
on $G$ in the non-exponential $d$-dependent regime
is a possible topic of further experimental work. 
Another subject of experimental investigation could be the
shot noise. Puzzling, although shot noise is a concept put forward 
in the context of vacuum tubes \cite{Schottky:18}, we are not aware 
of shot noise $\mathcal{S}_{sh}$ studies in vacuum nanogaps. 
Such studies may provide supplementary information.
Both $G$ and $V_t$ sample information contained in the $I-V$-characteristic
(expressed via the first power of the transmission coefficient $\mathcal{T}$), 
while $\mathcal{S}_{sh}$ also comprises information on the second power
$\mathcal{T}^2$; in addition, at relatively large 
biases $V\sim V_t$, $\mathcal{S}_{sh}$ samples information
on a broader energy range around the Fermi energy, 
as pointed out recently \cite{Baldea:2014a}.

To end, we have to admit that we are unable to unravel the origin of 
this mysterious ghost current that seems to be present even in a
standard vacuum nanogap. However speculative this idea 
might seem --- a fact of which we are fully aware---, 
we believe that the experimental support presented here
in favor of a presently unknown (``ghost'') transport channel
can represent a sufficient justification for considering the scenario proposed here
a working hypothesis, which deserves further 
investigations. Nanoelectronics could be the main beneficiary.
%%%%%%%%%%%%%%%%%%%%%%%%%%%%%%%%%%%%%%%%%%%%%%%%%%%%%%%%%%%%%%%%%%%%%%%%%%%
% \subsection*{Supporting Information} \ldots This material is available free of charge via the Internet at http://pubs.acs.org.
%%%%%%%%%%%%%%%%%%%%%%%%%%%%%%%%%%%%%%%%%%%%%%%%%%%%%%%%%%%%%%%%%%%%%%%%%%% 
\section*{Acknowledgment}
%%%%%%%%%%%%%%%%%%%%%%%%%%%%%%%%%%%%%%%%%%%%%%%%%%%%%%%%%%%%%%%%%%%%%%%%%%% 
The author thanks Prof.~Harold Zandvliet for providing him with
the data of his group prior to publication and numerous valuable
experimental details and comments.
Financial support provided by the Deu\-tsche For\-schungs\-ge\-mein\-schaft 
(grant BA 1799/2-1) is gratefully acknowledged.\\
% {\bf Electronic Supplementary Information {\esi}:} Supplementary material (computational details and tables) is available in the online version of this article at http://\ldots.
%%%%%%%%%%%%%%%%%%%%%%%%%%%%%%%%%%%%%%%%%%%%%%%%%%%%%%%%%%%%%%%%%%%%%%%%%%%%%%%%
\renewcommand\refname{Notes and references}
\providecommand*{\mcitethebibliography}{\thebibliography}
\csname @ifundefined\endcsname{endmcitethebibliography}
{\let\endmcitethebibliography\endthebibliography}{}


\begin{mcitethebibliography}{25}
\providecommand*{\natexlab}[1]{#1}
\providecommand*{\mciteSetBstSublistMode}[1]{}
\providecommand*{\mciteSetBstMaxWidthForm}[2]{}
\providecommand*{\mciteBstWouldAddEndPuncttrue}
  {\def\EndOfBibitem{\unskip.}}
\providecommand*{\mciteBstWouldAddEndPunctfalse}
  {\let\EndOfBibitem\relax}
\providecommand*{\mciteSetBstMidEndSepPunct}[3]{}
\providecommand*{\mciteSetBstSublistLabelBeginEnd}[3]{}
\providecommand*{\EndOfBibitem}{}
\mciteSetBstSublistMode{f}
\mciteSetBstMaxWidthForm{subitem}
{(\emph{\alph{mcitesubitemcount}})}
\mciteSetBstSublistLabelBeginEnd{\mcitemaxwidthsubitemform\space}
{\relax}{\relax}

\bibitem[Kushmerick(2009)]{kushmerick:09}
J.~Kushmerick, \emph{Nature}, 2009, \textbf{462}, 994--995\relax
\mciteBstWouldAddEndPuncttrue
\mciteSetBstMidEndSepPunct{\mcitedefaultmidpunct}
{\mcitedefaultendpunct}{\mcitedefaultseppunct}\relax
\EndOfBibitem
\bibitem[Song \emph{et~al.}(2011)Song, Reed, and Lee]{Reed:11}
H.~Song, M.~A. Reed and T.~Lee, \emph{Adv. Mater.}, 2011, \textbf{23},
  1583--1608\relax
\mciteBstWouldAddEndPuncttrue
\mciteSetBstMidEndSepPunct{\mcitedefaultmidpunct}
{\mcitedefaultendpunct}{\mcitedefaultseppunct}\relax
\EndOfBibitem
\bibitem[Beebe \emph{et~al.}(2006)Beebe, Kim, Gadzuk, Frisbie, and
  Kushmerick]{Beebe:06}
J.~M. Beebe, B.~Kim, J.~W. Gadzuk, C.~D. Frisbie and J.~G. Kushmerick,
  \emph{Phys. Rev. Lett.}, 2006, \textbf{97}, 026801\relax
\mciteBstWouldAddEndPuncttrue
\mciteSetBstMidEndSepPunct{\mcitedefaultmidpunct}
{\mcitedefaultendpunct}{\mcitedefaultseppunct}\relax
\EndOfBibitem
\bibitem[Sotthewes \emph{et~al.}()Sotthewes, Hellenthal, Kumar, and
  Zandvliet]{harold-letter}
K.~Sotthewes, C.~Hellenthal, A.~Kumar and H.~J.~M. Zandvliet, \emph{RSC Adv.,
  DOI 10.1039/C4RA04651J}\relax
\mciteBstWouldAddEndPuncttrue
\mciteSetBstMidEndSepPunct{\mcitedefaultmidpunct}
{\mcitedefaultendpunct}{\mcitedefaultseppunct}\relax
\EndOfBibitem
\bibitem[Choi \emph{et~al.}(2008)Choi, Kim, and Frisbie]{Choi:08}
S.~H. Choi, B.~Kim and C.~D. Frisbie, \emph{Science}, 2008, \textbf{320},
  1482--1486\relax
\mciteBstWouldAddEndPuncttrue
\mciteSetBstMidEndSepPunct{\mcitedefaultmidpunct}
{\mcitedefaultendpunct}{\mcitedefaultseppunct}\relax
\EndOfBibitem
\bibitem[Choi \emph{et~al.}(2010)Choi, Risko, Delgado, Kim, Bredas, and
  Frisbie]{Choi:10}
S.~H. Choi, C.~Risko, M.~C.~R. Delgado, B.~Kim, J.-L. Bredas and C.~D. Frisbie,
  \emph{J. Amer Chem. Soc.}, 2010, \textbf{132}, 4358 -- 4368\relax
\mciteBstWouldAddEndPuncttrue
\mciteSetBstMidEndSepPunct{\mcitedefaultmidpunct}
{\mcitedefaultendpunct}{\mcitedefaultseppunct}\relax
\EndOfBibitem
\bibitem[Hines \emph{et~al.}(2010)Hines, Diez-Perez, Hihath, Liu, Wang, Zhao,
  Zhou, M\"ullen, and Tao]{Tao:10}
T.~Hines, I.~Diez-Perez, J.~Hihath, H.~Liu, Z.-S. Wang, J.~Zhao, G.~Zhou,
  K.~M\"ullen and N.~Tao, \emph{J. Am. Chem. Soc.}, 2010, \textbf{132},
  11658--11664\relax
\mciteBstWouldAddEndPuncttrue
\mciteSetBstMidEndSepPunct{\mcitedefaultmidpunct}
{\mcitedefaultendpunct}{\mcitedefaultseppunct}\relax
\EndOfBibitem
\bibitem[Sommerfeld and Bethe(1933)]{Sommerfeld:33}
A.~Sommerfeld and H.~Bethe, \emph{Handbuch der Physik}, Julius-Springer-Verlag,
  Berlin, 1933, vol. 24 (2), p. 446\relax
\mciteBstWouldAddEndPuncttrue
\mciteSetBstMidEndSepPunct{\mcitedefaultmidpunct}
{\mcitedefaultendpunct}{\mcitedefaultseppunct}\relax
\EndOfBibitem
\bibitem[B\^aldea and K\"oppel(2012)]{Baldea:2012c}
I.~B\^aldea and H.~K\"oppel, \emph{Phys. Lett. A}, 2012, \textbf{376}, 1472 --
  1476\relax
\mciteBstWouldAddEndPuncttrue
\mciteSetBstMidEndSepPunct{\mcitedefaultmidpunct}
{\mcitedefaultendpunct}{\mcitedefaultseppunct}\relax
\EndOfBibitem
\bibitem[B\^aldea(2012)]{Baldea:2012e}
I.~B\^aldea, \emph{Europhys. Lett.}, 2012, \textbf{98}, 17010\relax
\mciteBstWouldAddEndPuncttrue
\mciteSetBstMidEndSepPunct{\mcitedefaultmidpunct}
{\mcitedefaultendpunct}{\mcitedefaultseppunct}\relax
\EndOfBibitem
\bibitem[B\^aldea and K\"oppel(2012)]{Baldea:2012f}
I.~B\^aldea and H.~K\"oppel, \emph{Phys. Stat. Solidi (b)}, 2012, \textbf{249},
  1791--1804\relax
\mciteBstWouldAddEndPuncttrue
\mciteSetBstMidEndSepPunct{\mcitedefaultmidpunct}
{\mcitedefaultendpunct}{\mcitedefaultseppunct}\relax
\EndOfBibitem
\bibitem[B\^aldea(2012)]{Baldea:2012b}
I.~B\^aldea, \emph{Chem. Phys.}, 2012, \textbf{400}, 65--71\relax
\mciteBstWouldAddEndPuncttrue
\mciteSetBstMidEndSepPunct{\mcitedefaultmidpunct}
{\mcitedefaultendpunct}{\mcitedefaultseppunct}\relax
\EndOfBibitem
\bibitem[Huisman \emph{et~al.}(2009)Huisman, Gu\'edon, van Wees, and van~der
  Molen]{Huisman:09}
E.~H. Huisman, C.~M. Gu\'edon, B.~J. van Wees and S.~J. van~der Molen,
  \emph{Nano Lett.}, 2009, \textbf{9}, 3909-- 3913\relax
\mciteBstWouldAddEndPuncttrue
\mciteSetBstMidEndSepPunct{\mcitedefaultmidpunct}
{\mcitedefaultendpunct}{\mcitedefaultseppunct}\relax
\EndOfBibitem
\bibitem[Trouwborst \emph{et~al.}(2011)Trouwborst, Martin, Smit, Gu\'edon,
  Baart, van~der Molen, and van Ruitenbeek]{Molen:11}
M.~L. Trouwborst, C.~A. Martin, R.~H.~M. Smit, C.~M. Gu\'edon, T.~A. Baart,
  S.~J. van~der Molen and J.~M. van Ruitenbeek, \emph{Nano Lett.}, 2011,
  \textbf{11}, 614--617\relax
\mciteBstWouldAddEndPuncttrue
\mciteSetBstMidEndSepPunct{\mcitedefaultmidpunct}
{\mcitedefaultendpunct}{\mcitedefaultseppunct}\relax
\EndOfBibitem
\bibitem[B\^aldea(2012)]{Baldea:2012h}
I.~B\^aldea, \emph{J. Phys. Chem. Solids}, 2012, \textbf{73}, 1151 --
  1153\relax
\mciteBstWouldAddEndPuncttrue
\mciteSetBstMidEndSepPunct{\mcitedefaultmidpunct}
{\mcitedefaultendpunct}{\mcitedefaultseppunct}\relax
\EndOfBibitem
\bibitem[Gundlach(1966)]{Gundlach:66}
K.~H. Gundlach, \emph{Solid-State Electronics}, 1966, \textbf{9}, 949 --
  957\relax
\mciteBstWouldAddEndPuncttrue
\mciteSetBstMidEndSepPunct{\mcitedefaultmidpunct}
{\mcitedefaultendpunct}{\mcitedefaultseppunct}\relax
\EndOfBibitem
\bibitem[B\^aldea(2013)]{Baldea:2013b}
I.~B\^aldea, \emph{Nanoscale}, 2013, \textbf{5}, 9222--9230\relax
\mciteBstWouldAddEndPuncttrue
\mciteSetBstMidEndSepPunct{\mcitedefaultmidpunct}
{\mcitedefaultendpunct}{\mcitedefaultseppunct}\relax
\EndOfBibitem
\bibitem[Zandvliet(private communication)]{harold-private}
H.~J.~W. Zandvliet, private communication\relax
\mciteBstWouldAddEndPuncttrue
\mciteSetBstMidEndSepPunct{\mcitedefaultmidpunct}
{\mcitedefaultendpunct}{\mcitedefaultseppunct}\relax
\EndOfBibitem
\bibitem[B\^aldea(2014)]{Baldea:2014a}
I.~B\^aldea, \emph{J. Phys. Chem. C}, 2014, \textbf{118}, 8676--8684\relax
\mciteBstWouldAddEndPuncttrue
\mciteSetBstMidEndSepPunct{\mcitedefaultmidpunct}
{\mcitedefaultendpunct}{\mcitedefaultseppunct}\relax
\EndOfBibitem
\bibitem[Guo \emph{et~al.}(2011)Guo, Hihath, Diez-P\'erez, and Tao]{Guo:11}
S.~Guo, J.~Hihath, I.~Diez-P\'erez and N.~Tao, \emph{J. Am. Chem. Soc.}, 2011,
  \textbf{133}, 19189--19197\relax
\mciteBstWouldAddEndPuncttrue
\mciteSetBstMidEndSepPunct{\mcitedefaultmidpunct}
{\mcitedefaultendpunct}{\mcitedefaultseppunct}\relax
\EndOfBibitem
\bibitem[B\^aldea(2012)]{Baldea:2012a}
I.~B\^aldea, \emph{Phys. Rev. B}, 2012, \textbf{85}, 035442\relax
\mciteBstWouldAddEndPuncttrue
\mciteSetBstMidEndSepPunct{\mcitedefaultmidpunct}
{\mcitedefaultendpunct}{\mcitedefaultseppunct}\relax
\EndOfBibitem
\bibitem[B\^aldea(2012)]{Baldea:2012g}
I.~B\^aldea, \emph{J. Am. Chem. Soc.}, 2012, \textbf{134}, 7958--7962\relax
\mciteBstWouldAddEndPuncttrue
\mciteSetBstMidEndSepPunct{\mcitedefaultmidpunct}
{\mcitedefaultendpunct}{\mcitedefaultseppunct}\relax
\EndOfBibitem
\bibitem[Wu \emph{et~al.}(2013)Wu, Bai, Sanvito, and Hou]{Hou:12}
K.~Wu, M.~Bai, S.~Sanvito and S.~Hou, \emph{Nanotechnology}, 2013, \textbf{24},
  025203\relax
\mciteBstWouldAddEndPuncttrue
\mciteSetBstMidEndSepPunct{\mcitedefaultmidpunct}
{\mcitedefaultendpunct}{\mcitedefaultseppunct}\relax
\EndOfBibitem
\bibitem[Wu \emph{et~al.}(2013)Wu, Bai, Sanvito, and Hou]{Hou:13}
K.~Wu, M.~Bai, S.~Sanvito and S.~Hou, \emph{J. Chem. Phys.}, 2013,
  \textbf{139}, 194703\relax
\mciteBstWouldAddEndPuncttrue
\mciteSetBstMidEndSepPunct{\mcitedefaultmidpunct}
{\mcitedefaultendpunct}{\mcitedefaultseppunct}\relax
\EndOfBibitem
\bibitem[Schottky(1918)]{Schottky:18}
W.~Schottky, \emph{Annalen der Physik}, 1918, \textbf{362}, 541--567\relax
\mciteBstWouldAddEndPuncttrue
\mciteSetBstMidEndSepPunct{\mcitedefaultmidpunct}
{\mcitedefaultendpunct}{\mcitedefaultseppunct}\relax
\EndOfBibitem
\end{mcitethebibliography}
\end{document}